# Frustrated Ferroelectricity in Niobate Pyrochlores

T. M. McQueen[1], B. Muegge[1], Q. Huang[2], K. Noble[1], H. W. Zandbergen[3] and R. J. Cava[1]

[1]Department of Chemistry, Princeton University Princeton NJ 08544
[2]NIST Center for Neutron Research, Gaithersburg Md 20899
[3]Department of Nanoscience, TU Delft, The Netherlands

**Abstract**

The crystal structures of the $A_2B_2O_{7-x}$ Niobium-based pyrochlores $Y_2(Nb_{0.86}Y_{0.14})_2O_{6.91}$, $CaYNb_2O_7$, and $Y_2NbTiO_7$ are reported, determined by powder neutron diffraction. These compounds represent the first observation of B-site displacements in the pyrochlore structure: the B-site ions are found to be displaced from the ideal pyrochlore positions, creating electric dipoles. The orientations of these dipoles are fully analogous to orientations of the magnetic moments in Ising spin based magnetically frustrated pyrochlores. Diffuse scattering in electron diffraction patterns shows that the displacements are only short range ordered, indicative of geometric frustration of the collective dielectric state of the materials. Comparison to the crystal structure of the $Nb^{5+}$ ($d^0$) pyrochlore $La_2ScNbO_7$ supports the prediction that charge singlets, driven by the tendency of Nb to form metal-metal bonds, are present in these pyrochlores. The observed lack of long-range order to these singlets suggests that $Nb^{4+}$-based pyrochlores represent the dielectric analogy to the geometric frustration of magnetic moments observed in rare earth pyrochlores.

Introduction

$A_2B_2O_7$ pyrochlores are of great recent interest as magnetic materials because their crystal structures, with both A-site and B-site metal sublattices in the geometry of corner sharing tetrahedra (Figure 1(a)), frustrate long range magnetic ordering.[1] As a consequence of crystal field effects, the Ising-like spins in some rare earth pyrochlores must point either directly into or directly out from the centers of the metal tetrahedra, leading to the observation of unexpected magnetic states such as "spin ice".[2, 3] Niobium oxides, by contrast, make many of the best known ferroelectrics and dielectrics and are virtually always non-magnetic, with the only exception being certain crystallographic shear structure compounds.[4-6] In this context, the recent proposal that non-magnetic behavior in a $Y^{3+}$-$Nb^{4+}$-O pyrochlore is due to the formation of spin singlets is unexpected.[7] This experimental work motivated a theoretical study[8] of a hypothetical $Y_2Nb_2O_7$ pyrochlore that led to the novel proposal that charge singlets, not spin singlets, are present in this compound. The proposed charge singlets would be formed by displacements of the $Nb^{4+}$ from their ideal pyrochlore B-site positions, energetically driven by niobium's propensity to form metal-metal bonds.[8] As a test of this novel proposal, we report here our determination of the crystal structures of three single phase $Nb^{4+}$-based pyrochlores, $Y_4Nb_3O_{12}$, $Y_2NbTiO_7$, and $CaYNb_2O_7$, by powder neutron diffraction and electron diffraction analysis. We observe displacements of the $Nb^{4+}$ ions from their ideal sites. These displacements result in electric dipoles that point either directly towards or directly away from the centers of the B-site sublattice tetrahedra (Figure 1(b)). Further, comparison to the structure of the $Nb^{5+}$ ($d^0$) based pyrochlore $La_2ScNbO_7$ shows that the displacements are likely to be electronically driven, as proposed in the charge singlet model. The electric dipole orientations that result from the displaced niobiums are exactly equivalent to the spin orientations seen in the spin ice magnetic pyrochlores. The displacements are not long-range ordered, frustrated by the lattice geometry, indicating that that these $Nb^{4+}$ pyrochlores are the dielectric equivalents of the Ising pyrochlore magnets: in other words, they are "dipole ices". The concept of dipole ice has recently been introduced as an important consequence of the lone-pair driven A-site displacements in $Bi_2Ti_2O_7$ and $Pb_2Sn_2O_6$ pyrochlores.[9, 10] The electronic origin of the dipole ice state in the niobate pyrochlores makes them a new, charge-driven manifestation of this new type of dielectric behavior.

Stoichiometric lanthanide niobate $Ln_2Nb^{4+}{}_2O_7$ (Ln = rare earth) pyrochlores, based solely on $Nb^{4+}$, are not known.[11] Our investigation of the chemistry of $Nb^{4+}$-based pyrochlores for the current study confirms their absence in equilibrium syntheses at one atmosphere pressure at temperatures up to 1625 °C. The pyrochlore in the Y-Nb-O system near a Y:Nb ratio of 1:1 is nonstoichiometric, and has the formula $Y_4Nb_3O_{12}$. All the Nb present is in the 4+ state, a $4d^1$ configuration that would lead to a spin ½ system if the electrons were localized, and the compound is analogous to the pyrochlore in the Y-Zr-O system, $Y_4Zr_3O_{12}$.[12] Here we report the crystal structure of this phase, which in a more familiar pyrochlore formula is written $Y_2(Y_{0.14}Nb_{0.84})_2O_{6.91}$; $Y^{3+}$ and $Nb^{4+}$ are mixed on the pyrochlore B-site and there are an appropriate number of charge-compensating oxygen vacancies. Stoichiometric niobium-based pyrochlores are, however, stable for electron counts corresponding to $Nb^{4.5+}$, [13, 14] and we here report the structures of two such pyrochlores, for the previously reported compound $CaYNb_2O_7$,[13, 14] and a new phase with the same electron count, $Y_2TiNbO_7$. To test whether displacements of the B-site atoms from their ideal sites in the $Nb^{4+}$-containing pyrochlores are electronically driven, we also determined the crystal structure of the $Nb^{5+}$-based pyrochlore $La_2ScNbO_7$.

**Experimental**

Starting materials for the syntheses were high purity $Ca_4Nb_2O_9$ (prepared in air at 1300° C from $CaCO_3$ (Mallinckrodt, 99.9%) and $Nb_2O_5$), $Y_2O_3$ (Alfa-Aesar, 99.9%), $Nb_2O_5$ (Aldrich, 99.9%), $NbO_2$ (Johnson-Matthey, 99.8%), $TiO_2$ (Aldrich, 99.9%), $La_2O_3$ (Alfa-Aesar, 99.99%), and $Sc_2O_3$ (Alfa-Aesar 99.999%). The starting materials were mixed in appropriate ratios, pressed into pellets, wrapped in Mo foil, and heated in a vacuum furnace back-filled with argon at 1625 °C for 32 hours, with one intermediate grinding and re-pelleting. The neutron diffraction patterns were taken at 298 K and 4 K at the NIST center for neutron research. A Cu (311) monochromator with a 90° take-off angle, $\lambda$ = 1.5404(2) Å, and an in-pile collimation of 15' were used. Data were collected over the two theta range 3–168° with a step size of 0.05°. The GSAS program suite was used for Rietveld structural refinement.[15] Scattering lengths (in *fm*) employed in the refinement were 7.75, 4.70, 8.24, 7.05, -3.44, 12.29, and 5.80 for Y, Ca, La, Nb, Ti, Sc, and O, respectively. To test for the correct space group, and the possibility of correlated Nb displacements over short distances, an electron diffraction study was performed with a Philips CM 200 electron microscope equipped with a field emission gun.

**Results**

We confirm that the only single phase pyrochlore in the Y:Nb:O system is found at an atomic ratio of 4:3:12, consistent with a previous report.[16] The room temperature neutron powder diffraction pattern for this phase is shown in Figure 2(a). The signature pyrochlore reflections (marked with arrows) are clearly visible, and the relative intensity of these peaks is similar to what is seen in other rare earth pyrochlores.[17] The neutron diffraction patterns for

$Y_2NbTiO_7$ (Figure 2(b)), and $CaYNb_2O_7$ (Figure 2(c)) are similarly clean, showing that these compositions are single phase.

The neutron powder diffraction data for all three compounds are fit well by cubic pyrochlore structures with the space group Fd-3m (#227). In the normal cubic pyrochlore structure, the A-site ions are in special position 16c (0,0,0), the oxygens are in 48f (x,1/8,1/8) and 8a (1/8,1/8,1/8), and the B-site ions are found in the special positions 16d (1/2,1/2,1/2). This space group allows for the possibility of statistical displacements of the B-site ions from their ideal coordinates either to positions of the type (1/2+δ, 1/2+δ, 1/2+δ), forming two equivalent positions with 1/2 occupancy displaced along <111> and <-1 -1 -1>, or to positions of type (1/2+δ, 1/2+δ, 1/2), forming six equivalent positions with 1/6 occupancy displaced along <1,1,0>. For all three $Nb^{4+}$-based pyrochlores we studied, the structural refinements showed that when the B-site ions are constrained to be in the ideal (1/2,1/2,1/2) site, the thermal vibration parameters are unusually large, an indication that the B-site ions are displaced from their ideal positions. Data for undisplaced models for all three compounds are given in Table II. In our test of the different Nb displacement models, the only one to successfully describe the data is one in which Nb atoms are displaced along <111> directions (Figure 2(d)). For all three compounds, the magnitudes of the displacements are highly significant: ranging from 23 standard deviations to 57 standard deviations. In all cases, the improvements of the $\chi^2$ values are significant, and the thermal vibration parameters assume a more normal value when the B-site ions are placed in displaced positions.

With the displacements of the B-site ions established through the refinements in space group Fd-3m, the question arises as to whether the displacements occur in a disordered or ordered fashion. In Fd-3m the displacements are statistical, and not ordered. We found that the neutron diffraction data are also fit well by pyrochlore structures in which the B-site displacements are ordered – with a "two in, two out" ordered B-site displacement, as shown in Figure 2(e). This ordered displacement model, where two Nb atoms are displaced inward and two are displaced outward in every tetrahedron (consistent with Pauling's "ice rules"[18, 19]) is still cubic, but has the lower symmetry space group F-43m (#216). The neutron powder diffraction fit statistics comparing the statistical and ordered displacement models are shown in Table I for $CaYNb_2O_7$ (#227-II and #216(a) respectively). To facilitate the most direct comparison, the disordered displacement model was also run in the lower symmetry F-43m space group (#216(b), the coordinates in brackets represent the results from the #227-II fit translated into F-43m), where two symmetry inequivalent <1,1,1> type displacement directions were allowed for the B-sites, and their relative occupancies were refined. All three fits have comparable $\chi^2$ values (#227-II: 1.199, #216(a): 1.212, #216(b): 1.264) Both models also contain highly significant (approximately 40 standard deviations) displacements of the B-site atoms from the ideal pyrochlore position with reasonable thermal vibration parameters. Thus the neutron powder data clearly shows that the niobium atoms are displaced from their ideal positions in the pyrochlore structure, as predicted. Whether the displacements are ordered or disordered, however (i.e. the choice of space group), cannot be determined solely from the neutron powder data as our sensitivity is not sufficiently high.

High dynamic-range electron diffraction was performed to resolve this ambiguity. (hk0) reflections with $h + k = 2n$ are allowed in space group F-43m but are not permitted in space group Fd-3m, for which only $h + k = 4n$ is allowed. Since niobium is a strong scatterer, the strongest in the material $CaYNb_2O_7$, for example, a niobium position displacement of the magnitude obtained from the neutron data would result in a detectable peak at the (600) position in electron diffraction if the space group is F-43m. Figures 3a,b, and c show the (001) diffraction zone of each pyrochlore. The <hk0> family of reflections with $h+k=2n$ (some locations marked with small circles) are absent in all three samples. This shows that a disordered B-site displacement model (Figure 1(b)), with the ideal pyrochlore space group Fd-3m, is the correct average structure for all three compounds. This conclusion is determined to high sensitivity due to the high dynamic range of the electron diffraction patterns.

The refined structure parameters, based on the disordered B-site displacement model from the neutron powder diffraction data for all three $Nb^{4+}$ containing compounds are presented in Table II, and the room temperature fits are displayed in Figure 2. All the structural parameters for the compounds are of the ideal pyrochlore type, with the exception of the B-site position. For $CaYNb_2O_7$, Ca and Y are randomly mixed on the A-site, and the B-site is fully occupied by $Nb^{4.5+}$. For the nonstoichiometric pyrochlore $Y_4Nb_3O_{12}$, the pyrochlore formula is $Y_2(Nb_{0.86}Y_{0.14})_2O_{6.91}$, i.e. Y fully occupies the A-site, Y and Nb are mixed on the B-site, and the oxygen sublattice has a small number of vacancies. For $Y_2NbTiO_7$, Y occupies the A site and Nb and Ti randomly occupy the B-site. The $\chi^2$ values for fits to the neutron diffraction data taken at 298 K are all excellent ($CaYNb_2O_7$: 1.199, $Y_4Nb_3O_{12}$: 1.270, $Y_2NbTiO_7$: 1.018), and the thermal vibration parameters are reasonable and well defined. The refined oxygen content for $Y_4Nb_3O_{12}$, which in the pyrochlore formula is $Y_2(Nb_{0.86}Y_{0.14})_2O_{6.91}$, is found to be 6.95(5), consistent with the expected formula. Neutron diffraction measurements at 4 K on $CaYNb_2O_7$ show that the lattice constants and thermal vibration parameters decrease with decreasing temperature, but the displacements of the Nb atoms from the ideal B-site position remain relatively unchanged, indicating that the atomic displacements are fully developed at room temperature. Recalling that the x value for the ideal position is 0.5000, in $Y_4Nb_3O_{12}$ the refined B-site position is 0.5114(2), in $Y_2NbTiO_7$ it is 0.5116(5), and in the compound with single-atom B-site occupancy, $CaYNb_2O_7$, it is 0.5081(2).

Displacements of $Nb^{5+}$ ions from the centers of their coordinating oxygen octahedra are well known in perovskite-like structures, often giving rise to ferroelectric behavior. For comparison, therefore, in the pyrochlore case, we also collected neutron diffraction data for the related $Nb^{5+}$-based pyrochlore compound $La_2ScNbO_7$. The observed diffraction data and the intensities calculated from a fit to an ideal pyrochlore model in space group Fd-3m are shown in Figure 4. This compound contains $Nb^{5+}$ and $Sc^{3+}$ randomly distributed on the B-site, both $d^0$ ions, and, unlike the three reduced niobates studied here, does not show any significant displacement of the atoms from the ideal B position. As shown in Table II, the ideal and displaced models for $La_2ScNbO_7$ give equivalently good fit statistics. Additionally, the freely refined position of the B-site in $La_2ScNbO_7$ is 0.4994(9), which is, within one standard deviation and to good precision, equal to the ideal value of 0.5. This suggests that the origin of the displacements in the $Nb^{4+}$-based pyrochlores is likely the proposed formation of charge singlets by $Nb^{4+}$ ($4d^1$) ions.[8] Although numerous $Nb^{5+}$-based pyrochlore systems are known (e.g. $Bi_2InNb^{5+}O_7$, [20] and $A_2Nb^{5+}{}_2O_7$ (A = Ca, Cd, Hg) [11]) that have been studied structurally, to the best of our knowledge this is the first observation of B-site displacements in a pyrochlore, consistent with an uncommon driving force for their occurrence. For $d^0$ $M^{4+}$-based pyrochlores (e.g. $La_2Zr^{4+}{}_2O_7$, [21] $Bi_2Ti^{4+}{}_2O_7$, [9] and $Ln_2Ti^{4+}{}_2O_7$ [17]) no B-site displacements have ever been reported, again supporting an uncommon driving force for the $d^1$ $Nb^{4+}$ cases studied here.

Frustration of the B-site displacements is supported by electron diffraction, which shows short range ordering of the atomic shifts. As seen in Figure 3(a,b,c), the <hk0> ($h + k = 2n$) reflections are absent, but there is diffuse scattering near those positions for all three (although for $Y_2NbTiO_7$ the diffuse scattering is very weak), suggestive of short-range ordering. Electron diffraction patterns for each compound taken slightly tilted off the (001) zone are shown in Figures 3(d, e, and f). In addition to the primary reflections, there is significant diffuse scattering, more than in the (001) zone, further evidence of some type of short range order. Although the diffuse scattering has a complex shape in three dimensions that we have not analyzed in detail, some general features are evident. First, the diffuse scattering is centered around the expected locations of the reflections that would indicate an ordered "two in, two out" displacement of B-sites. Second, the magnitude of the diffuse scattering correlates well with the composition of the B-site: it is greatest when the B-site is stoichiometric, in $(Ca_{0.5}Y_{0.5})_2Nb_2O_7$ (Figure 3(d)), less when 1/7 of the Nb on the B-site is replaced ($Y_2(Nb_{0.86}Y_{0.14})_2O_{6.91}$, Figure 3(e)), and almost undetectable when 1/2 of the Nb is replaced ($Y_2(Nb_{0.5}Ti_{0.5})_2O_7$, Figure 3(f)). Thus compositional disordering of the B-site seems to disrupt the short range ordering of the displacements even as the magnitude of the displacements increases (see Table II), supporting the conclusion that the diffuse scattering comes from short range ordered displacements of the B-site ions. Contributions other than correlations in the B-site displacements, e.g. from short range order on the B-site, cannot be strictly ruled out, but the fact that the diffuse scattering is strongest for the case where the B-site is fully occupied by Nb argues against this. The same can be said for oxygen vacancy ordering, as such vacancies are not present in $CaYNb_2O_7$. The position of the diffuse scattering changes with compositional disorder: it is far from the expected positions of reflections for an ordered "two in, two out" distortion when the B-site is stoichiometric (large circle, Figure 3(d)), but much closer when the B-site is slightly disordered (large circle, Figure 3(e)). This is consistent with the displacements of the B-site cations being geometrically frustrated, prevented from forming a long range ordering of "two in, two out" distortions when the B-site is close to stoichiometric. While the displacements are not as frustrated when there is more B-site disorder, as indicated by the largest displacement in $Y_2(Nb_{0.5}Ti_{0.5})_2O_7$ (see Table II), the same disorder prevents long range ordering of the displacements, as observed by the negligible diffuse scattering in $Y_2(Nb_{0.5}Ti_{0.5})_2O_7$ (large circle, Figure 3(f)).

In conclusion, our detailed powder neutron and electron diffraction measurements reveal that the three reduced Nb pyrochlores, $Y_2(Nb_{0.86}Y_{0.14})_2O_{6.91}$ ($Y_4Nb_3O_{12}$), $CaYNb_2O_7$ and $Y_2NbTiO_7$ have average structures that display displacements of the transition metal ions on the B-site sublattice, along <111> directions. There appears to be short range ordering of the B-site displacements. This, combined with the relative magnitude of the displacements, suggests geometric frustration of the electric dipoles, which are formed by Nb displacement along the <111> directions. Comparison to the fully $d^0$ chemically analogous pyrochlore $La_2ScNbO_7$ shows that these displacements must be present for electronic reasons, and is consistent with the predicted charge singlets. The geometry of the pyrochlore lattice prevents formation of a single ordered ground state made from ordered displacements of this type, as evidenced by the complex diffuse scattering from short range ordering. Exploring further consequences of the presence of the frustrated ferromagnetic, "dipole ice" state in these compounds would be of interest in future studies, particularly to elucidate the structure of the short range ordering of Nb displacements and compare it to the predicted contraction of the Nb tetrahedra. [8]

**Acknowledgements**

T. M. McQueen gratefully acknowledges support by the national science foundation graduate research fellowship program. This work was done under NSF grant DMR06-20234. Certain commercial materials and equipment are identified in this report to describe the subject adequately. Such identification does not imply recommendation or endorsement by the NIST, nor does it imply that the materials and equipment identified are necessarily the best available for the purpose.

**Table I**. Structural parameters for the CaYNb$_2$O$_7$ pyrochlore at 298 K obtained using Fd-3m (#227) and F-43m (#216) space groups. For #227-II, **A**: 16c (0 0 0), **B**: 32e(x x x), and **O(1)**: 48f (x, 1/8, 1/8), and **O(2)**: 8a (1/8, 1/8, 1/8). For #216, **A**: 16e (x x x), **B**: 16e (x x x), **B'**: 16e (x x x), **O(1)**: 24g (x, ¼, ¼), **O(1)'**: 24f (x 0 0), **O(2)**: 4a (0 0 0), **O(2)'**: 4c (¼, ¼, ¼).

| Parameter | | #227-II | | #216(a) | #216(b) |
|---|---|---|---|---|---|
| | a (Å) | 10.3161(2) | | 10.3160(2) | 10.3159(2) |
| **A** | x | | | 1/8 | 1/8 |
| | B (Å$^2$) | 1.04(3) | | 1.03(3) | 0.98(3) |
| | n (Ca/Y) | 0.5/0.5 | | 0.5/0.5 | 0.5/0.5 |
| **B** | x | 0.5081(2) | [0.6331*] | 0.6332(2) | 0.6309(2) |
| | B (Å$^2$) | 1.00(3) | | 0.99(4) | 1.24(3) |
| | n | 0.5 | | 0.5 | 1 |
| **B'** | x | | | 0.6168(2) | |
| | B (Å$^2$) | | | 0.99(2) | |
| | n | | | 0.5 | |
| **O(1)** | x | 0.4220(1) | [0.5470*] | 0.5448(7) | 0.5465(5) |
| | B$_{11}$ (Å$^2$) | 1.45(3) | | 1.41(4) | 1.38(3) |
| | B$_{22}$=B$_{33}$ (Å$^2$) | 1.05(2) | | 1.05(2) | 1.04(2) |
| | n | 1 | | 1 | 1 |
| **O(1)'** | x | | [0.2970*] | 0.2992(7) | 0.2966(5) |
| | B$_{11}$ (Å$^2$) | | | 1.41(4) | 1.38(3) |
| | B$_{22}$=B$_{33}$ (Å$^2$) | | | 1.05(2) | 1.04(2) |
| | n | | | 1 | 1 |
| **O(2)** | B (Å$^2$) | 1.03(3) | | 0.86(5) | 1.01(5) |
| **O(2)'** | B (Å$^2$) | | | 0.86(5) | 1.01(5) |
| | R$_P$ % | 3.75 | | 3.77 | 3.81 |
| | R$_{wp}$ % | 4.74 | | 4.76 | 4.86 |
| | $\chi^2$ | 1.199 | | 1.212 | 1.264 |

* x ± 1/8.

**Table II**. Structural parameters for the pyrochlores CaYNb$_2$O$_7$, Y$_2$(Nb$_{0.86}$Y$_{0.14}$)$_2$O$_{6.91}$, and Y$_2$NbTiO$_7$ at 298 K (first line), and, for CaYNb$_2$O$_7$, 4 K (second line). The non-reduced Nb pyrochlore, La$_2$ScNbO$_7$, is included for comparison, as are refined parameters without B-site displacements. Space group *Fd-3m* (#227). Atomic positions: **A**: 16*c* (0 0 0), **O(1)**: 48*f* (x, 1/8, 1/8), **O(2)**: 8*a* (1/8, 1/8, 1/8), for displaced: **B**: 32e (x x x), and for ideal: **B**: 16d (½ ½ ½).

| Parameter | | CaYNb$_2$O$_7$ | | Y$_2$(Nb$_{0.86}$Y$_{0.14}$)$_2$O$_{6.91}$ | | Y$_2$NbTiO$_7$ | | La$_2$ScNbO$_7$ | |
|---|---|---|---|---|---|---|---|---|---|
| | | Displaced | Ideal | Displaced | Ideal | Displaced | Ideal | Displaced | Ideal |
| | a (Å) | 10.3161(2) 10.3046(3) | 10.3159(2) | 10.3281(2) | 10.3283(3) | 10.1807(1) | 10.1807(2) | 10.6742(1) | 10.6742(1) |
| **A** | B (Å$^2$) | 1.04(3) 0.84(6) | 1.01(3) | 1.34(4) | 1.21(2) | 0.81(2) | 0.82(2) | 1.05(2) | 1.05(2) |
| | n | 0.5/0.5 0.5/0.5 | 0.5/0.5 | 1 | 1 | 1 | 1 | 1 | 1 |
| **B** | x | 0.5081(2) 0.5086(4) | | 0.5114(2) | | 0.5116(5) | | 0.4994(9) | |
| | B (Å$^2$) | 1.00(3) 0.85(7) | 1.50(3) | 1.29(7) | 2.06(4) | 1.5(2) | 2.6(1) | 0.86(2) | 0.81(2) |
| | n | 0.5 0.5 | 1 | 0.43/0.07 | 0.86/0.14 | 0.25/0.25 | 0.5/0.5 | 0.25/0.25 | 0.5/0.5 |
| **O(1)** | x | 0.4220(1) 0.4220(2) | 0.4211(1) | 0.4077(2) | 0.4077(2) | 0.4155(1) | 0.4156(6) | 0.4231(1) | 0.4231(1) |
| | $B_{11}$ (Å$^2$) | 1.45(3) 1.41(7) | 1.57(4) | 4.0(1) | 2.79(8) | 1.44(4) | 1.38(4) | 1.59(3) | 1.61(3) |
| | $B_{22}=B_{33}$ (Å$^2$) | 1.05(2) 0.78(4) | 1.19(3) | 2.55(7) | 1.97(6) | 0.79(2) | 0.75(2) | 0.93(2) | 0.94(2) |
| | n | 1 1 | 1 | 0.992(8) | 0.947(8) | 1 | 1 | 1 | 1 |
| **O(2)** | B (Å$^2$) | 1.03(3) 1.18(6) | 1.22(4) | 0.95(9) | 0.95(2) | 0.41(4) | 0.47(3) | 0.60(4) | 0.56(4) |
| | R$_P$ % | 3.75 6.68 | 3.94 | 4.26 | 5.58 | 3.54 | 3.61 | 4.51 | 4.46 |
| | R$_{wp}$ % | 4.74 8.38 | 5.07 | 5.11 | 7.02 | 4.44 | 4.51 | 5.69 | 5.64 |
| | $\chi^2$ | 1.199 0.965 | 1.370 | 1.270 | 1.603 | 1.018 | 1.060 | 1.464 | 1.440 |

**Figure 1**. (a) The ideal cubic pyrochlore structure consists of two interpenetrating corner-sharing tetrahedral lattices (oxygen atoms not shown for clarity). In the reduced niobates studied, the "B" site atoms are displaced along the <111> directions, as shown in (b). The displacements along these directions make the electric dipoles in the reduced niobate pyrochlores the dielectric analogy of spin ice compounds. In contrast, the B-site lattice of the $Nb^{5+}$-only pyrochlore $La_2ScNbO_7$, remains undistorted (as in (a)).

**Figure 2**. (a) Shows the neutron powder diffraction data for $Y_4Nb_3O_{12}$ at 298 K, fit to a disordered B-site displacement model. The quality of the fit is excellent, with no unexplained peaks and no systematic trend in the residuals. Some pyrochlore superreflections are marked with arrows. Similar fits of (b) $CaYNb_2O_7$ and (c) $Y_2NbTiO_7$ are also shown. (d) Shows a tetrahedral unit for disordered displacement of Nb atoms along the <111> directions, where each Nb atom in the tetrahedron displaces either into or away from the center of the tetrahedron. By contrast, (e) shows the model for an ordered displacement of Nb atoms along the <111> directions, with "two in, two out" in every tetrahedron.

**Figure 3**. High dynamic range (~$10^4$) electron diffraction images of the (001) zone for (a) $CaYNb_2O_7$, (b) $Y_4Nb_3O_{12}$, and (c) $Y_2NbTiO_7$. The <hk0> ($h + k = 2n$) family of reflections (marked with small circles), expected for the F-43m model, are not present in any of the three compounds, though there is significant diffuse scattering near these reflections. The data shows that the space group is Fd-3m and not F-43m. Electron diffraction images taken slightly tilted off the (00l) zone illustrates the complex nature of the diffuse scattering centered near the formally absent reflections (marked with large circles): (d) $CaYNb_2O_7$ shows intense diffuse scattering;

(e) $Y_4Nb_3O_{12}$ also shows diffuse scattering off zone, but it is much weaker than in (d); (f) $Y_2NbTiO_7$ shows almost no diffuse scattering, indicating that the short-range ordering is basically lost. This is consistent with frustration of electric dipoles on the B-site.

**Figure 4**. Powder neutron diffraction pattern of the $La_2ScNbO_7$ pyrochlore, showing an excellent fit to the observed data that indicates that there is no displacement of the B-site ions in this material.

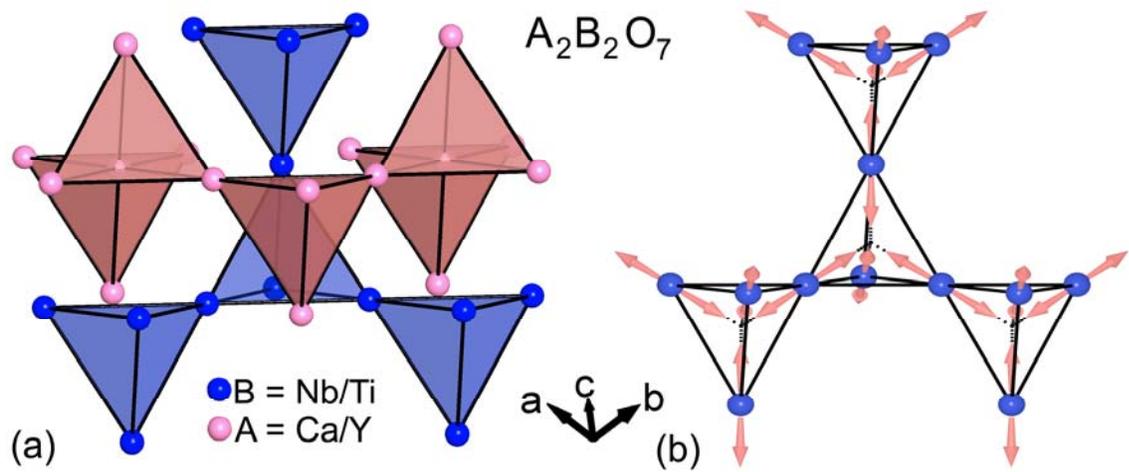

**Figure 1**

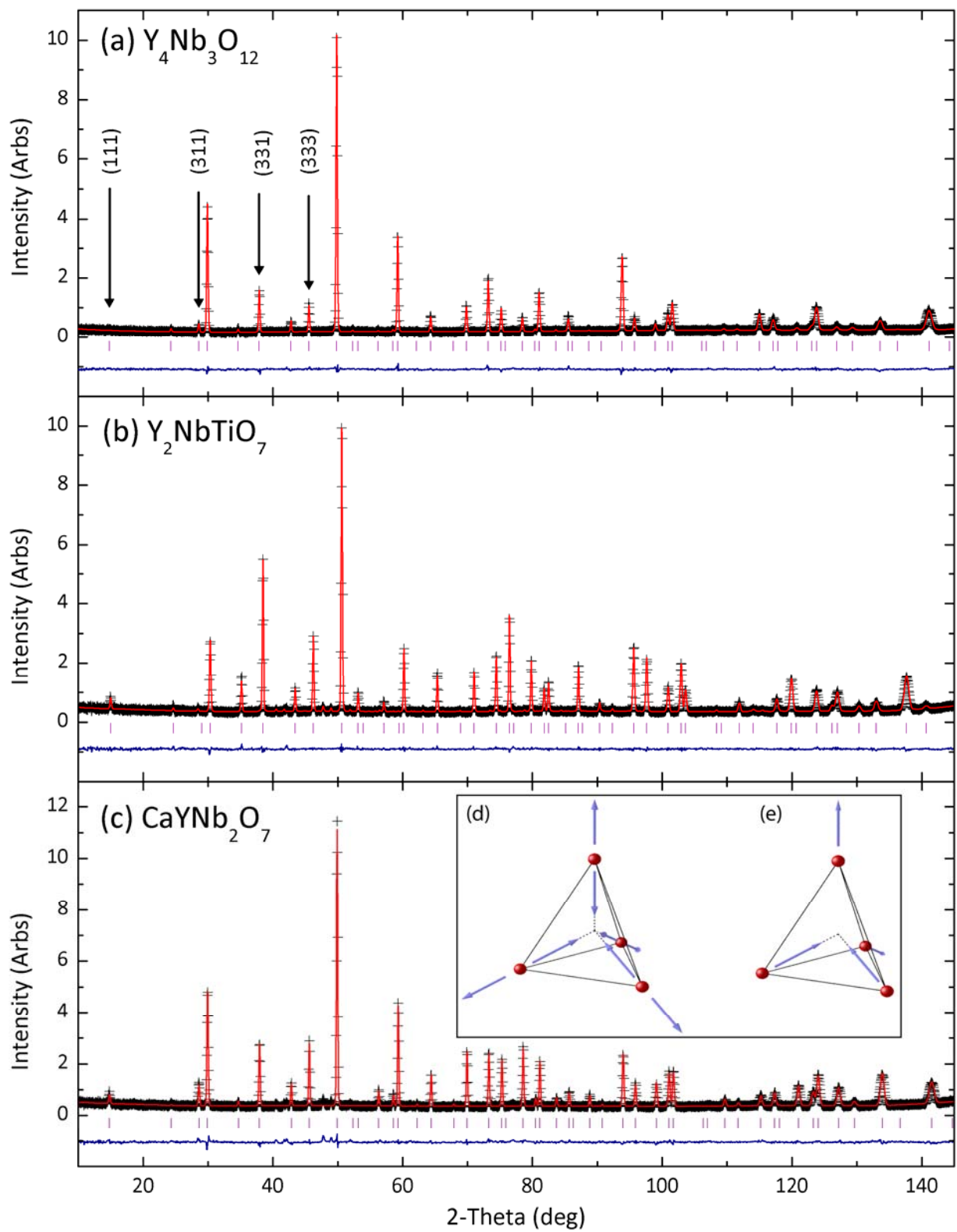

**Figure 2**

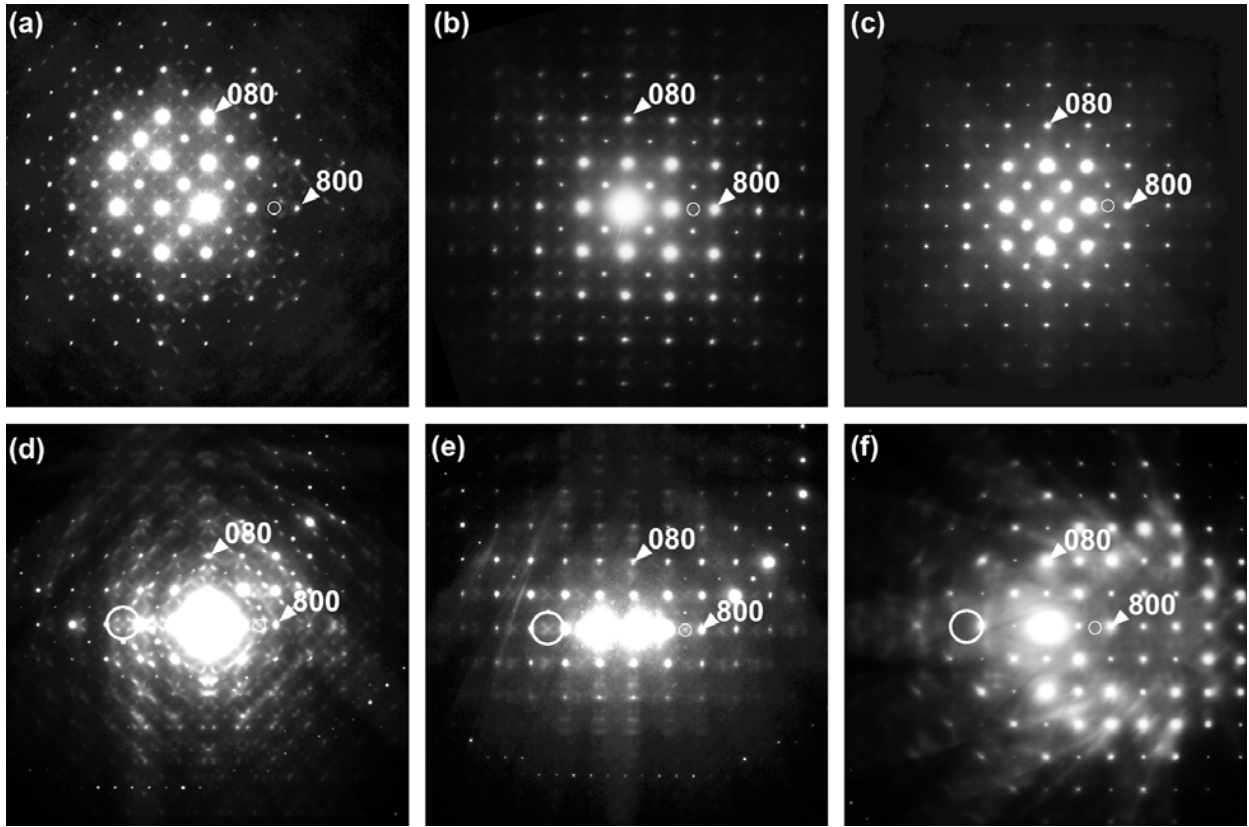

**Figure 3**

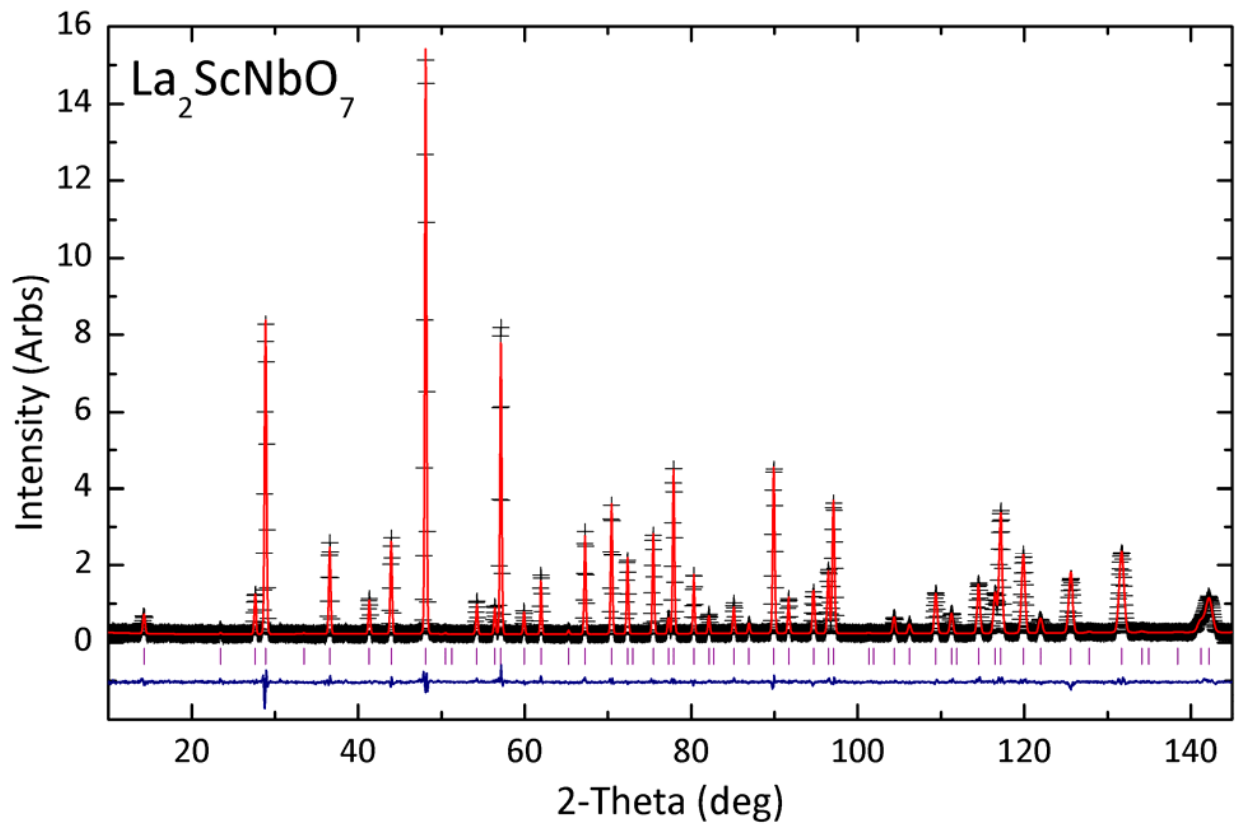

**Figure 4**